\documentclass[prd,twocolumn,preprintnumbers]{revtex4-1}

\usepackage{amsmath}
\usepackage{epsfig}
\usepackage{graphicx}
\usepackage{color}
\usepackage[normalem]{ulem}
 \usepackage{url}
\usepackage[breaklinks, plainpages=false, colorlinks=true, anchorcolor=cyan, linkcolor=red, citecolor=cyan, urlcolor=magenta, bookmarks=false]{hyperref}

\usepackage[caption=false]{subfig}

\begin{document}
\renewcommand{\thefigure}{\arabic{figure}}
\setcounter{figure}{0}

 \def\I{{\rm i}}
 \def\E{{\rm e}}
 \def\D{{\rm d}}

\bibliographystyle{apsrev}

\title{Black Hole Hunting with LISA}

\author{Neil J. Cornish}
\affiliation{eXtreme Gravity Institute, Department of Physics, Montana State University, Bozeman, Montana 59717, USA}

\author{Kevin Shuman}
\affiliation{eXtreme Gravity Institute, Department of Physics, Montana State University, Bozeman, Montana 59717, USA}

\begin{abstract} 
The Laser Interferometer Space Antenna (LISA) will be able to detect massive black hole mergers throughout the visible Universe.  These observations will provide unique information about black hole formation and growth, and the role black holes play in galaxy evolution. Here we develop several key building blocks for detecting and characterizing black hole binary mergers with LISA, including fast heterodyned likelihood evaluations, and efficient stochastic search techniques. 
\end{abstract}

\maketitle

\section{Introduction}

The first detection by LIGO of gravitational waves from a binary black hole merger~\cite{Abbott:2016blz} has been followed by dozens of additional detections~\cite{LIGOScientific:2018mvr} that are revealing insights into stellar evolution and black hole formation~\cite{LIGOScientific:2018jsj}. In the next decade, the launch of the Laser Interferometer Space Antenna (LISA)~\cite{Audley:2017drz,Baker:2019nia} will allow for similar studies of much more massive black holes, potentially allowing us to unravel the interplay between massive black hole growth and galaxy evolution.

The detection and characterization of binary black hole mergers was investigated two decades ago~\cite{Cornish:2006ry,Cornish:2006dt,Cornish:2006ms, Cornish:2007jv,Rover:2007iq,Babak:2008bu,Porter:2008kn,Petiteau:2009pi,Gair:2009cx} as part of the planning for an earlier incarnation of the LISA mission. In the intervening years there have been several improvements in both the modeling of the signals and the techniques used to detect and characterize the signals. The most significant changes are that the signal models now include inspiral, merger and ringdown, as opposed to just the inspiral, and a range of techniques have been developed that greatly speed up the calculation of the likelihood function, which plays a central role in the analysis. These techniques have recently been used in a study of Bayesian parameter estimation for LISA observation of massive black hole binaries, with an emphasis on the impact of including higher harmonics in the signal model~\cite{Marsat:2020rtl,Katz:2020hku}. 

The goal of this work is to develop several key building blocks for detecting massive black hole mergers with LISA and inferring their physical properties. The effort is part of the LISA Data Challenge~\cite{LDC}, a successor to the original series of Mock LISA Data Challenges~\cite{Arnaud:2006gm,Babak:2008aa,Babak:2009cj}, where simulated LISA data is used as a playground for developing analysis algorithms that can be used once the mission is operational.  The LISA Data Challenges are following a staged development, starting with relatively simple data sets and progressively building in additional realism. In the first round of the new challenges, dubbed {\em Radler}, the data sets are broken out by source type. For massive black holes the simulated data set contains a single binary merger in uninterrupted stationary, Gaussian noise. The next challenge, {\em Sangria}, will include simulated data with multiple black hole binaries. Future challenges will add gaps, non-stationary and non-Gaussian noise, and will include multiple signals of different types, as well as increasing the complexity of the simulated signals. The techniques described here are sufficient to handle the {\em Radler} and {\em Sangria} Challenges, and will serve as a foundation for the development of the more advanced techniques needed to handle more realistic data sets that will ultimately form part of the global solution that simultaneously models thousands of overlapping signals of different types. Our approach is similar to that in Ref.~\cite{Marsat:2020rtl}, but with a greater emphasis on the initial search. Another key difference is that our analysis accommodates instrument noise, while the analysis in Ref.~\cite{Marsat:2020rtl} is limited to noise-free data. The GPU accelerated likelihood approached used in Ref.~\cite{Katz:2020hku} is able to account for instrument noise.

To avoid getting bogged down in details, most of the technical aspects of the analysis, such as the instrument response function, noise spectra {\it etc} are relegated to appendices. Geometric units with $G=c=1$ are used throughout.

\section{The Quarry}

The black hole mergers we are considering have total masses between $10^5 \, M_\odot$ and $10^8 \, M_\odot$. Lower mass systems, including the stellar origin black holes detected by LIGO and Virgo, will require slightly different search techniques due to their longer duration. For the signal model we use a phenomenological model (PhenomD), which describes the dominant harmonic of a quasi-circular binary with spins aligned with the orbital angular momentum~\cite{Husa:2015iqa,Khan:2015jqa}. The model attaches an augmented post-Newtonian inspiral to a parameterized merger and ringdown, and the model is calibrated against a suite of numerical relativity simulations. Including additional harmonic content in the signal model will have little impact on the search strategy, however allowing for mis-aligned spins and orbital precession would require modifications to the search.

\begin{figure}[htp]
\includegraphics[width=0.5\textwidth]{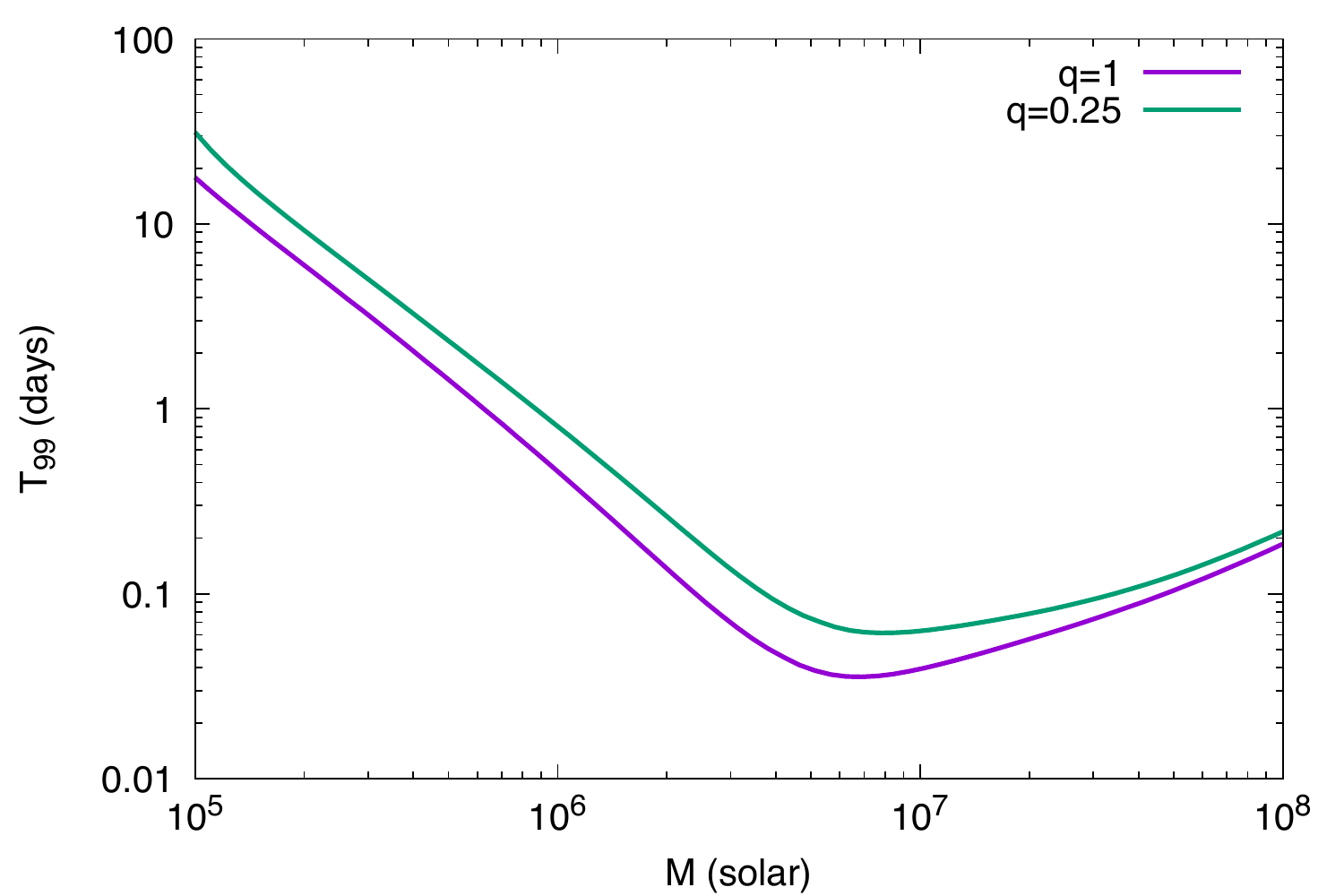} 
\caption{\label{fig:SNR99} The time interval before merger to accumulate 99\% of the signal-to-noise ratio squared, $T_{99}$, as a function of the detector-frame total mass for two different mass ratios. For binaries with total mass above $10^5 M_\odot$ the bulk of the signal-to-noise is accumulated in less than one month.}
\end{figure}

The search strategy is guided by the fact that systems in the mass range being considered only spend a short amount of time in the LISA band relative to the mission duration (years) and orbital modulation time scale (months). It is conventional to define ``time in band'' as the time to merger from some fiducial frequency, but a more meaningful measure is the time until merger during which a large fraction of the signal-to-noise ratio, or SNR, is accumulated. Since the Bayes factor between signal and noise scales as the signal-to-noise ratio squared, we define the time in-band, $T_{99}$, to be the time before merger taken to accumulate 99\% of the ${\rm SNR}^2$. Figure~\ref{fig:SNR99} shows the time in-band as a function of the detector-frame total mass $M$ for two different mass ratios $q=m_2/m_1$. The source was simulated with ecliptic co-latitude $\theta=\pi/3$ and longitude $\phi=0$, with polarization angle $\psi=\pi/3$ and inclination angle $\iota = \pi/3$. The choice of source location and orientation is largely irrelevant here since the time in-band is so short - the LISA antenna is roughly constant on durations shorter than one month. Here we are using the LISA mission configuration and noise model described in the LDC manual~\cite{LDCmanual}. Details of the noise model are given in Appendix~\ref{a:noise}.

\section{The Hunt}

Because the time in-band is short for systems with total mass greater than $10^5 \, M_\odot$, an efficient search strategy is to analyze shorter stretches of data. During the mission this could be done on a rolling basis, with the data segment being advanced day-by-day as the data arrives in an effort to provide low-latency alerts to aid searches for electromagnetic counterparts. Note though that there is little hope of providing advanced warning of a merger for systems with moderate mass ratios and detector frame total masses above $3\times 10^5  \, M_\odot$.

The short duration of the signals allows us to ignore the antenna response in the first stage of the search, cutting the search space from eleven dimensions (two masses, two spins, merger time and phase, distance, sky location and orbital orientation) to just four - the two masses $m_1,m_2$ and the two dimensionless spins $\chi_1, \chi_2$. The waveform amplitude, phase and merger time are maximized over analytically using the methods described in Section 8 of the LIGO Data Analysis guide~\cite{LIGOScientific:2019hgc}. In the LISA setting we analyze two channels of data, the signal-orthogonal $A$ and $E$ time-delay-interferometry (TDI) channels. The merger time maximization is performed simultaneously for both channels, while the amplitude and phase maximization is performed individually in each channel. The search over masses and spins could be performed using a LIGO-style template bank, but we prefer to use a stochastic search that is a variant of the Markov Chain Monte Carlo (MCMC) method we use for parameter estimation.

The PhenomD~\cite{Husa:2015iqa,Khan:2015jqa} waveform code provides the frequency domain amplitude ${\cal A}(f)$ and phase $\Phi(f)$ for the gravitational wave signal $h(f) = {\cal A}(f)e^{i \Phi(f)}$. To convert this to the fractional-frequency TDI response used in the LDC data sets we have to multiply the amplitude by a factor of $8 (f/f_*) \sin(f/f_*)$, where $f_* = 1/{2 \pi L} \simeq 19.1$ mHz is the transfer frequency and $L$ is the arm-length. The factor of $2 \sin(f/f_*)$ accounts for the time-delay interferometry, while the factor of $4 f/f_*$ accounts for the fractional frequency response. The time delay interferometry also introduces a phase shift of $\pi/2$ and a time shift of $L$, but these are taken care of by the analytic maximization.

\begin{figure}[htp]
\includegraphics[width=0.5\textwidth]{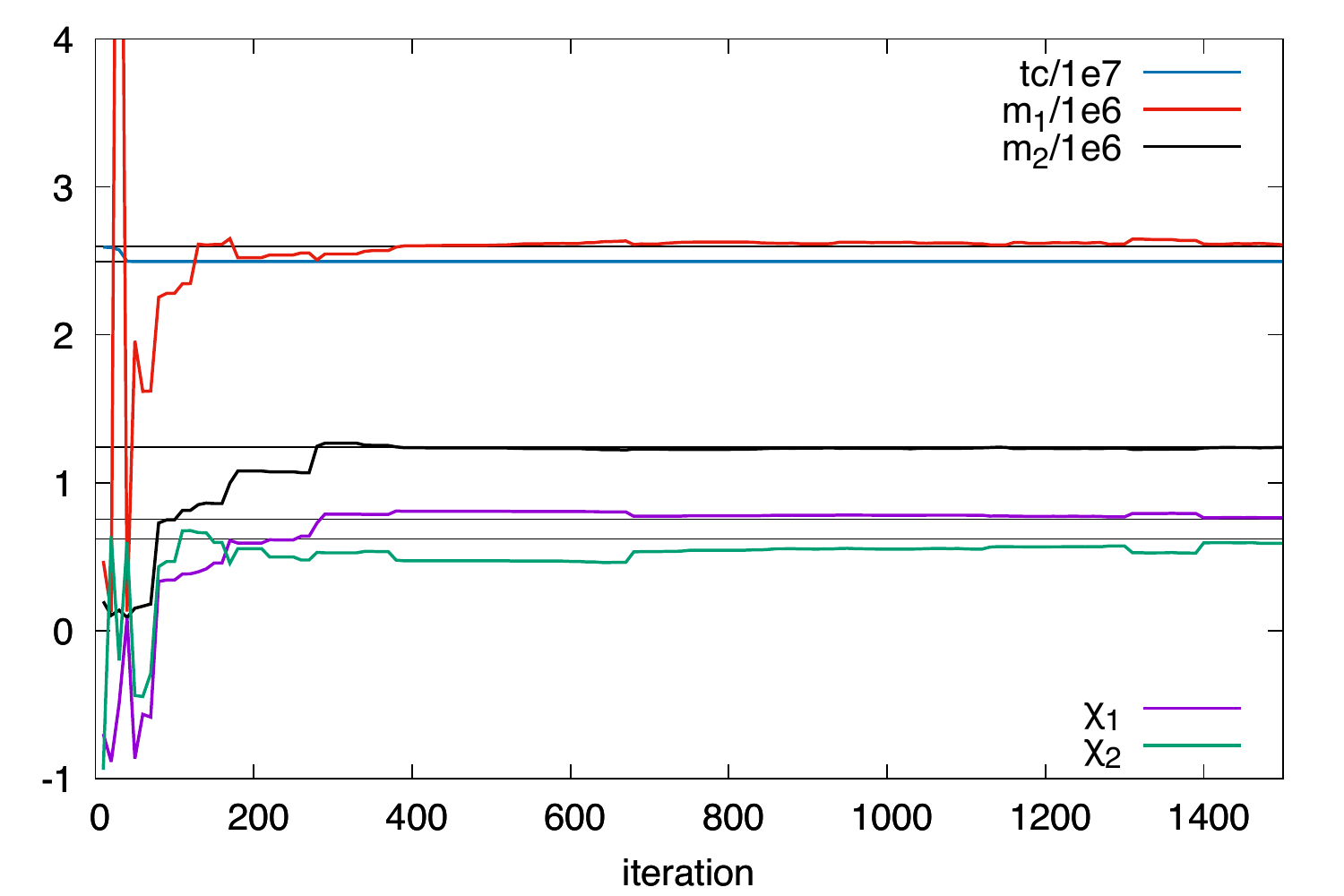} 
\caption{\label{fig:search} Trace plots of the cold chain from the stochastic search. The horizontal solid lines indicate the true parameter values. The search rapidly locks onto the signal.}
\end{figure}

The full $2^{22} \times 10$ seconds of the LDC massive black hole data set was divided into 16 chunks, each roughly a month in duration ($T_{\rm seg} = 2621440$ seconds). Each chunk was searched using a variant of the replica exchange MCMC~\cite{PhysRevLett.57.2607} that we use for parameter estimation. The search used a total of twelve chains, geometrically spaced in ``temperature'' by a factor of 1.5. The parameters of the cold chain were cloned to the hottest chain every 100 iterations. A mixture of proposals were used, including uniform draws from the prior range for each parameter, and draws along eigendirections of the Fisher information matrix, scaled by the inverse square root of the eigenvalues. The Fisher matrix was computed using the masses and spins, in addition to the merger time and phase. The merger time and phase were included in the Fisher matrix calculation, even though they were maximized over in the likelihood, since there are covariances between them and the masses and spins. Leaving out the merger time and phase in the Fisher matrix results in jump proposals that are inefficient.
The maximization over merger time $t_c$ in the calculation of the likelihood was restricted to be at most $\pm T_{\rm seg}/8.0$ so as not to go too far outside the frequency range of the reference signal, which was set to be from $f(T_{\rm start})$ to $ f(T_{\rm start}+T_{\rm seg})$ where $T_{\rm start}$ is the start time of the segment (the mapping between time and frequency is given in equation \ref{tf}). The prior for the merger time was set to be $t_c \in [T_{\rm start}, T_{\rm start}+2T_{\rm seg}]$, so that the merger could occur in the current segment or the next segment over, allowing for the possibility of picking up the signal prior to merger. The stochastic search is not Markovian (reversible) since the likelihood is maximized and the proposal densities are not included in the the Metropolis-Hastings jump acceptance probability.

Figure~\ref{fig:search} shows a trace plot of the merger time, masses and spins from the cold chain in the stochastic search of data chunk 10 of 16. As expected, the search rapidly locked onto the merger time, followed by the masses and then the spins. Searches of the other 15 segments yielded no additional significant candidate signals with ${\rm SNR} > 8$. The search of each segment takes less than four minutes using a quad-core 2.9 GHz Macbook Pro. The 12 chains were run in parallel using 
OpenMP. 

\begin{figure}[htp]
\includegraphics[width=0.5\textwidth]{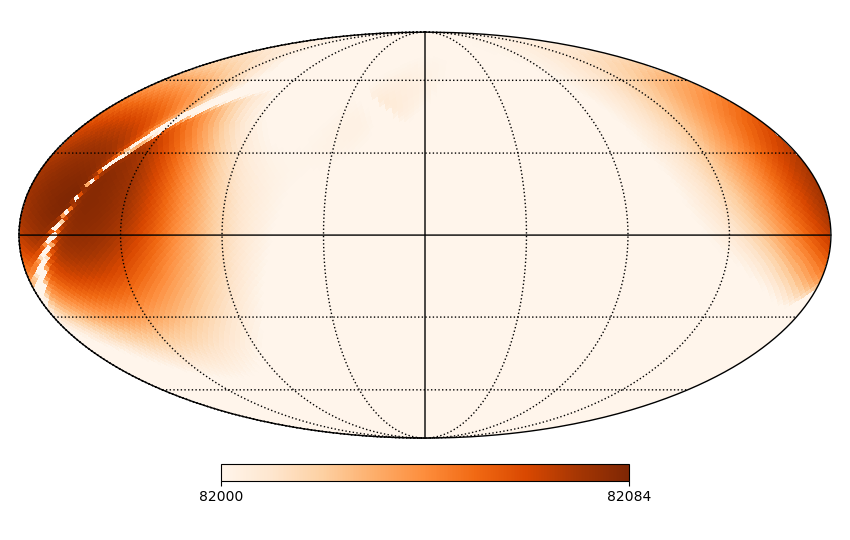} 
\caption{\label{fig:sky} A sky map showing the F-statistic likelihood computed using the values for the masses and spins found during the first phase of the search.}
\end{figure}

The initial stage of the search provides a starting solution for the masses, spins and the detector frame merger time. The next stage of the search finds a good starting solution for the sky location $(\theta,\phi)$, luminosity distance $D_L$, Barycenter merger time $t_c$, merger phase $\phi_c$, and orbital orientation (expressed in terms of the polarization angle $\psi$ and inclination angle $\iota$). The intrinsic parameters (masses and spins) are held fixed, while the other parameters are explored using a stochastic search algorithm. For the second stage of the search we need to apply the full instrument response to the PhenomD templates. The response is computed directly in the frequency domain using the method described in Appendix~\ref{a:response}. The likelihood is computed using the F-statistic~\cite{Jaranowski:1998qm} maximization described in Ref.~\cite{Cornish:2006ms}, which uses a set of four filters found by setting $\iota=\pi/2$ and (i) $(\phi_c,\psi) = (0,0)$; (ii) $(\phi_c,\psi) = (\pi/2,\pi/4)$; (iii)  $(\phi_c,\psi) = (3\pi/4,0)$;  (iv)  $(\phi_c,\psi) = (\pi/4,\pi/4)$; in the full response. The F-statistic maximizes the likelihood over the merger phase, luminosity distance, polarization angle and inclination angle, thus reducing the search to be over the sky location and Barycenter merger time. The Barycenter and detector frame merger times are related by $t_c^{\rm B} = t_c^{\rm D} + \hat{k}\cdot{\bf x}_0(t_c)$, where $ \hat{k}$ is the direction of propagation of the gravitational wave and ${\bf x}_0(t)$ is the center of the LISA constellation. For a given sky location for the source, this mapping can be used to estimate the Barycenter merger time. We allow the Barycenter merger time to vary a little from this value to account for the time delays introduced by the full instrument response. The F-statistic based search typically locks onto the true sky location in less than a hundred iterations.

The same search technique can be used on data sets containing multiple black hole mergers. The maximum likelihood solution from the previous pass is subtracted from the data and the search repeated. This process is repeated until no additional significant signals are found. The collection of maximum likelihood solutions from the search can be used as a starting point for more refined parameter estimation, or even better, can be turned into proposal distributions for performing the full multi-source global fit.

\section{Dressing Out}

With the rapid search phase complete and one or more sources identified, the next step is to refine the estimates for the source parameters. In reality this will be done while simultaneously inferring the parameters of many sources, including hundreds of massive black holes, tens of thousands of ultra-compact galactic binaries, hundreds of extreme mass ratio inspirals, and dozens of stellar origin black hole binaries. Additionally it will be necessary to model the instrument noise and residual galactic signal, which accounting for gaps in the data and other real-world complications.

Here we start with the simpler problem of inferring the parameters of a single massive black hole binary merger in gap-free data, with stationary, Gaussian noise with a known power spectrum. We apply Bayesian inference to compute the posterior distribution for the source parameters. Our method of choice is the replica exchange (parallel tempered) Markov Chain Monte Carlo (PTMCMC) algorithm. Over the years we have developed a standard recipe~\cite{2019SAAS...48....1C} for implementing PTMCMCs that uses a combination of local and global proposal distributions, and we adopt that approach here, while also adding a new ingredient - maximized jumps - that significantly improves the sampling. The individual chains are advanced using the Metropolis-Hastings algorithm, whereby a chain is advanced from parameters $x$ to parameters $y$ with acceptance probability
\begin{equation}
H(y | x)= {\rm min}\left( 1, \frac{ p( d|  y) p(y) q( x |  y)}{  p( d|  x) p( x) q(y |  x)} \right)\, .
\end{equation}
Here $p( d|  x)$ is the likelihood of observing data $d$ given model parameters $x$, $p(x)$ is the prior distribution for parameters x, and $q(y|x)$ is the proposal density for drawing a new set of parameters $y$ given the current set of parameters $x$.

One drawback of stochastic algorithms such as PTMCMCs is that they require large numbers of likelihood evaluations. Using the fast frequency domain technique described in Appendix~\ref{a:response}, applied to the full {\em Radler} data set, each likelihood evaluation takes roughly one second on a single 2.6 GHz CPU core. A variety of techniques can be used to speed up the likelihood evaluation, including reduced order models~\cite{Field_2011}, reduced-order quadratures~\cite{Canizares_2015} and computational approaches such as GPU acceleration~\cite{Katz:2020hku}. Here we use a different technique~\cite{Cornish:2010kf} that allows us to compute the likelihood without having to generate any waveforms aside from the one reference waveform that is used to heterodyne the data. The heterodyne likelihood has been rediscovered and used in LIGO/Virgo data analysis, though there the technique has been called ``relative binning''~\cite{Zackay:2018qdy}. In the current application, the heterodyned likelihood takes $\sim 1$ ms to compute - a factor of
one thousand times faster than direct evaluation with the (already fast) frequency domain waveforms. The method used to compute the heterodyned likelihood is described in Appendix~\ref{a:hlike}. 

\subsection{Priors}

We assumed uniform priors on all the parameters. The detector frame individual masses were taken to be uniform in the range $m_1,m_2 \in [5\times 10^4 M_\odot, 10^8 M_\odot]$. The dimensionless spins were taken to be uniform in the range $\chi_1,\chi_2 \in [-1,1]$. The luminosity distance was taken to be uniform in the range $D_L=[0.1 {\rm Gpc}, 400 {\rm Gpc}]$. The merger time $t_c$ was taken to be uniform in the range $[0,2 T_{\rm obs}]$, where $T_{\rm obs}$ is the observation time. The cosine of the ecliptic colatitude, $\cos\theta$, and the cosine of the inclination, $\cos\iota$, were taken to be uniform in the range $[-1,1]$. The orbital phase at merger $\phi_c$, and the polarization angle $\psi$ were taken to be uniform in the range $[0,\pi]$. The ecliptic longitude, $\phi$, was taken to be uniform in the range $[0,2\pi]$.

\subsection{Proposals}

We used a PTMCMC with a geometrically spaced temperature ladder, with inverse temperatures scaling as $\beta_i = \alpha^{-i}$. Samples are recorded from the chain with $\beta_0=1$. A total of $N_c=16$ sixteen chains were used, with each chain running on a separate computational core. The geometric temperature spacing $\alpha > 1$ was set so as to give an effective signal to noise ratio of ${\rm SNR}_{\rm eff} = {\rm SNR} / \beta_{N_c - 1} = 5$. 
More efficient sampling could probably be achieved using an adaptive temperature spacing. We did check that all the chains remained ``connected'', that is, that the exchange rate between neighboring chains remained above zero throughout the simulation and for all temperatures.

A mixture of proposal distributions were used to advance the chains. As per our standard recipe~\cite{2019SAAS...48....1C}, the mix included local and global proposals, with the choice of proposal at each iteration drawn randomly. To the standard mix we also added a new technique that incorporates maximization over parameters in a way that maintains detailed balance (reversibility) in the chains.

\begin{figure}[htp]
\includegraphics[width=0.5\textwidth]{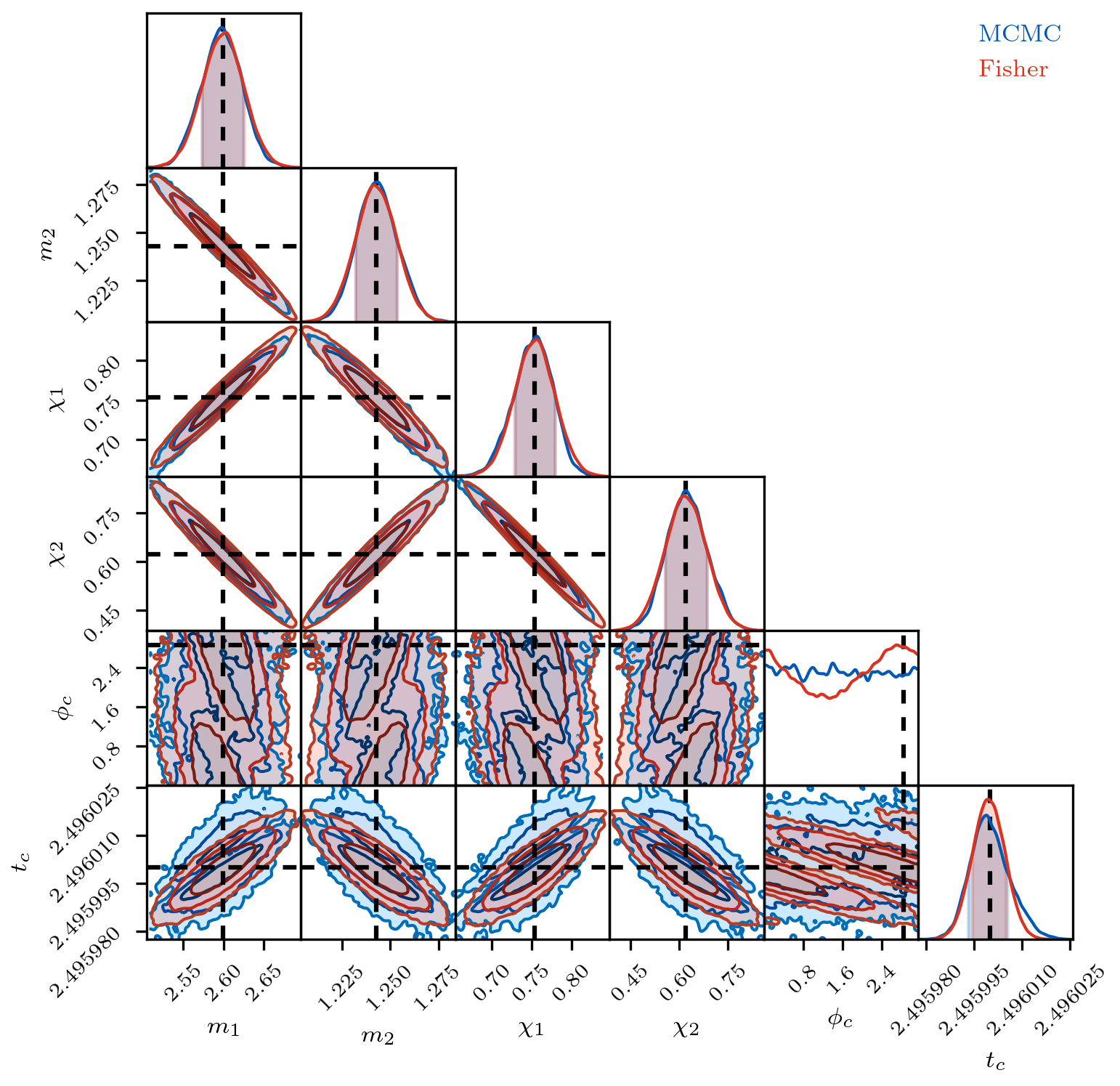} 
\caption{\label{fig:comp_in} Corner plot comparing the MCMC (blue) and Fisher matrix (red) estimates for the posterior distribution for the masses, spins, merger phase
and merger time. The true values are indicated by dashed lines. The two estimates agree well for the masses and spins, but less so for the merger time and phase. }
\end{figure}

\begin{figure}[htp]
\includegraphics[width=0.5\textwidth]{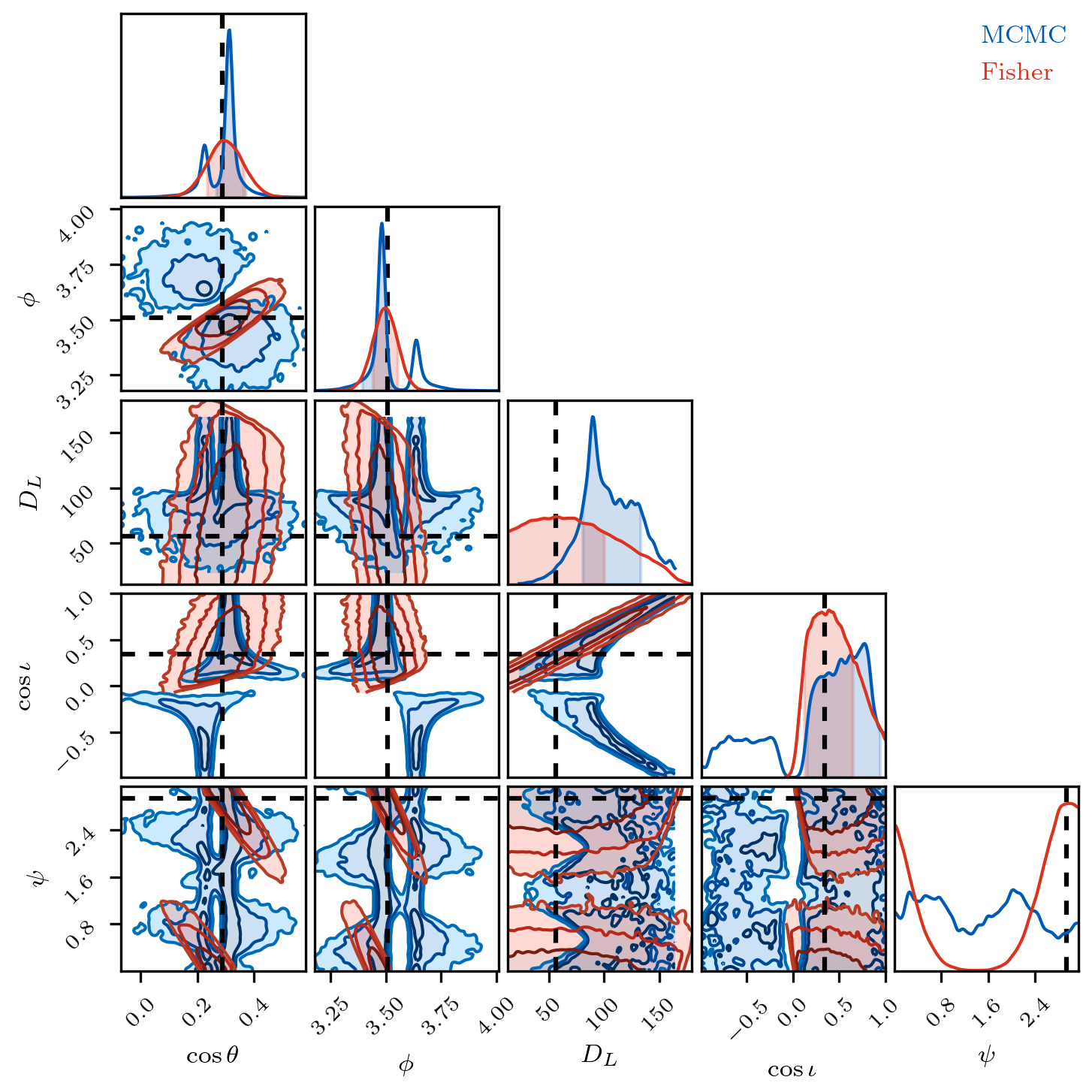} 
\caption{\label{fig:comp_ex} Corner plot comparing the MCMC (blue) and Fisher matrix (red) estimates for the posterior distribution for the sky location, luminosity distance, inclination and polarization. The true values are indicated by dashed lines.  The Fisher matrix provides a poor approximation to the MCMC derived posterior distribution
for these extrinsic parameters.}
\end{figure}

The ultimate proposal density would be posterior distribution itself, but lacking that, we instead use two local approximations to the posterior: proposals derived from previous chain samples and proposals that use the quadratic Fisher information matrix approximation to the likelihood.  A variant of the differential evolution approach~\cite{ter2006markov} is used to propose jumps based on previously collected samples. A running history of past samples is collected at each temperature level. 
Jumps are proposed from the current position to a new location found by adding the vector connecting two points drawn randomly from the history. The proposal can be shown to be asymptotically Markovian~\cite{ter2006markov}. Heuristically, the asymptotic reversibility can be understood from the observation that as the samples accumulate, the chain history approaches the stationary target distribution. Differential evolution is very effective at exploring strong parameter degeneracies, as the vectors connecting past samples tend to line up along the degenerate directions. The Fisher information matrix provides a quadratic approximation to the likelihood, which in turn is a good approximation to the posterior distribution so long as curvature of the prior is less than the curvature of the likelihood. The Fisher matrix can be
computed using a fast spline integration method~\cite{Cornish:2010kf}. Writing the signal as $h(f) = {\cal A}(f) e^{i \Phi(f)}$, the Fisher matrix $\Gamma_{ij} = (h_{,i} | h_{,j})$ can be expressed as
\begin{equation}\label{fish}
\Gamma_{ij} = 4 \sum_{I=A,E} \int \frac{{\cal A}^I_{,i}{\cal A}^I_{,j}+{{\cal A}^I}^2 \Phi^I_{,i}\Phi^I_{,j}}{S^I_n(f)} \, df \, .
\end{equation}
Here $S^I_n(f)$ is the noise spectral density in the $I^{\rm th}$ data channel, and the sum is over the $A,E$ TDI channels. All of the terms appearing in (\ref{fish}) vary slowly in frequency and can be evaluated on a coarse grid using the spline integration method described in Appendix~\ref{a:hlike}. The eigenvalues and eigenvectors of the Fisher matrix are computed and used to propose jumps by first randomly selecting an eigendirection, then drawing the jump size from a normal distribution with variance equal to the inverse of the corresponding eigenvalue. The effectiveness of these proposals is predicated on the Fisher matrix providing a reasonable approximation to the posterior distribution. Figures~\ref{fig:comp_in} and \ref{fig:comp_ex} compare the Fisher matrix approximation to the posterior to the MCMC derived posterior distributions for the noiseless {\em Radler} data set. We see that the Fisher matrix provides a good approximation for the parameters shown in  Figure~\ref{fig:comp_in} that enter directly into the gravitational wave phase, but the approximation is poor for the extrinsic parameters, such as sky location, luminosity distance and inclination angle shown in Figures~\ref{fig:comp_ex}. In part, the poor showing for the extrinsic parameters is because the distributions are multi-modal whereas the Fisher matrix approximation is mono-modal. The agreement is better in practice since the Fisher information matrices are updated as the simulation progresses, so all the modes get covered. On the other hand, the agreement seen in Figure~\ref{fig:comp_in} is deceiving, as the likelihoods computed using the Fisher matrix approximation are often very different from those computed using the full likelihood, mostly due to inaccuracies in the merger time and merger phase. These inaccuracies were found to severely limit the acceptance rate for jumps along certain eigendirections. Two strategies were used to improve the acceptance of the Fisher matrix based proposals. The first was to break the Fisher matrix into blocks and sometimes just propose jumps in the subset of parameters that are well approximated, the second was to use the likelihood maximization procedure described below.

\begin{figure}[htp]
\includegraphics[width=0.5\textwidth]{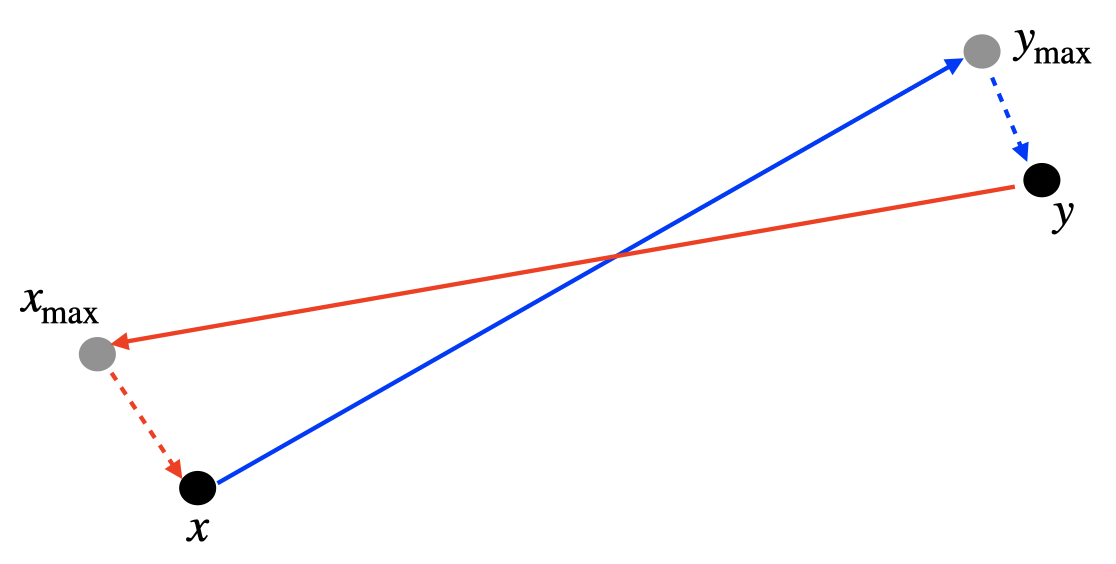} 
\caption{\label{fig:max} An illustration of the maximized jump proposal technique. A jump from the current location $x$ to a new location $y$ proceeds in two steps. First, a new location is proposed and parameters such as merger time and phase are analytically maximized over, yielding the new point $y_{\rm max}$. Next, a jump is drawn from a normal distribution from $y_{\rm max}$ to $y$. The proposal densities for the forward and reverse jumps are included in the Metropolis-Hastings ratio to ensure detailed balance.}
\end{figure}

The posterior distributions encountered in gravitational wave astronomy are often multi-modal. To fully explore all the modes, our approach~\cite{2019SAAS...48....1C} is to use global proposals combined with replica exchange (parallel tempering). Here we use a global proposal based on the F-statistic likelihood. A similar approach has been used for ultra compact galactic binaries~\cite{Littenberg:2020bxy}, but here the implementation is different. Rather than pre-computing a collection of F-statistic maps as was done in Refs~\cite{Littenberg:2020bxy, Becsy:2019dim}, here we compute the F-statistic likelihood on-demand, and use it as part of a maximized likelihood proposal. 

The maximized proposal technique is a general method that can be used with any form of likelihood maximization. In the current context we applied the technique to the F-statistic likelihood and also to likelihoods that are maximized with respect to time offset, overall phase and amplitude. The method is illustrated in Figure~\ref{fig:max}. The first step is to draw a new set of parameters from some distribution. For example, when using the F-statistic, a new sky location is drawn while holding the masses and spins fixed. The remaining parameters, $\{ t_c, \phi_c, \psi, \iota, D_L\}$ are analytically maximized using the F-statistic. The Barycenter merger time is adjusted to keep the detector frame merger time fixed. The new sky location is either drawn from the prior, or from a wide normal distribution centered on the current sky location. When coupled with the Fisher matrix proposal, the maximization is performed on $\{ t_c, \phi_c, D_L\}$. If uncorrected for, the maximization over parameters would violate detailed balance and bias the posterior distribution. To restore detailed balance a second step is added to the proposal: the Fisher information matrix is computed at $y_{\rm max}$ for the subset of parameters that are maximized over. A second jump $\Delta y$ is drawn from a  normal distribution with covariance matrix equal to the inverse of the Fisher matrix yielding the proposed point $y=y_{\rm max}+\Delta y$. The full proposal density $q(y|x)$ is then given by $q(y|x) = q(\Delta y | y_{\rm max})
 q(y_{\rm max} |x)$. The proposal density for the reverse move, $q(x|y) = q(\Delta x | x_{\rm max})  q(x_{\rm max} |y)$ is computed after finding the point $x_{\rm max}$ by maximizing the likelihood using the fixed parameters at $x$.
 
 The maximized proposal technique dramatically improves the mixing of the chains and the discovery of secondary posterior modes. For example, even without using dedicated proposals that exploit the symmetries of the LISA instrument response, maximized jumps using uniform draws on the sky location were able to quickly find all the secondary modes.

\subsection{Results}

\begin{figure}[htp]
\includegraphics[width=0.5\textwidth]{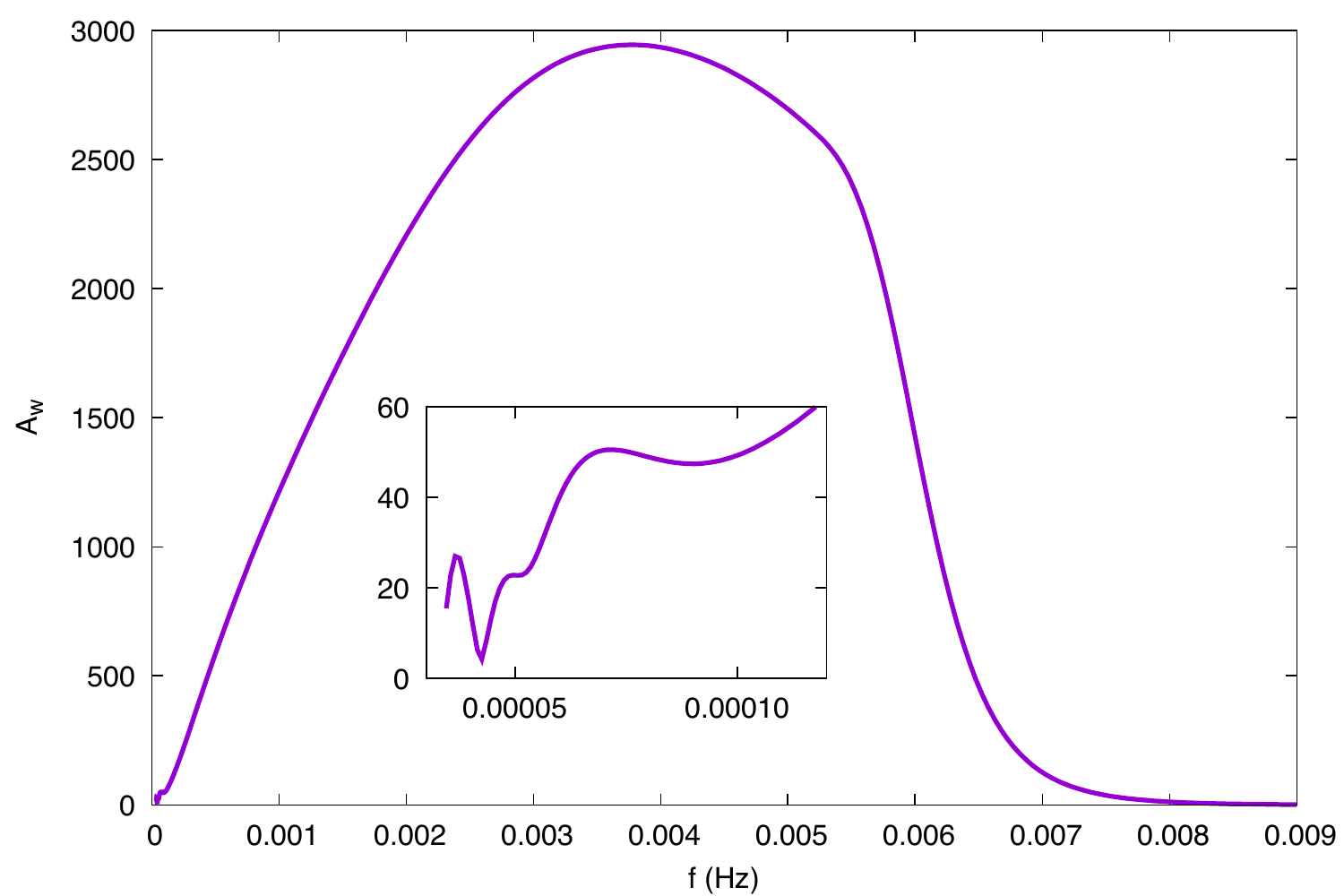} 
\caption{\label{fig:Awhite} The whitened signal amplitude in the TDI $A$ channel for the {\em Radler} black hole binary system. The inset highlights the amplitude modulation due to the LISA orbital motion.}
\end{figure}

The black hole binary system used in the {\em Radler} simulation has detector frame masses $m_1=2.599137 \times 10^6 \, M_{\odot}$, $m_2 = 1.242860 \times 10^6  \, M_{\odot}$ and dimesionless spins $\chi_1 = 0.75348$ and $\chi_2 = 0.62159$. The system was placed at a luminosity distance of $D_L = 56.006$ Gpc, corresponding to a redshift of $z=5.7309$ for the assumed cosmological model. The source frame masses are a much more modest $m_1^{s} = 3.8615 \times 10^5 \, M_{\odot}$, $m_2^s = 1.8465 \times 10^5  \, M_{\odot}$.  The relatively high detector frame total mass for this system results in it becoming detectable less than a day before merger: the signal reaches ${\rm SNR}=10$ just 11 hours prior to merger. The modulation of the amplitude and phase of the signal due to the LISA orbit is essentially irrelevant since very little SNR is accumulated prior to the last few hours before merger. Figure~\ref{fig:Awhite} shows the whitened signal amplitude for this system in the TDI $A$ channel, with an inset showing the amplitude modulation that occurs at low frequencies where the signal is undetectable. Consequently, the signal is effectively a short duration burst, and most of the directional information comes from differences in the time of arrival of the signal at each spacecraft, much like the situation for bursts from cosmic string cusps and kinks~\cite{Key:2008tt,Cohen:2010xd} or generic short duration bursts~\cite{Robson:2018jly}.

Unless data is downloaded from the LISA constellation every few hours or so, it is highly unlikely that systems such as this one will be detected prior to merger. If data were available in advance, the sky localization would be poor. To investigate this possibility, we smoothly truncated the time domain data using a cosine window of the form
\begin{equation}
W(t) = \left\{\begin{array}{lr}
        1 & t \leq T_{\rm cut}-\Delta T\\
         0 & t > T_{\rm cut} \\
        \frac{1}{2}\left({1-\cos\left( \frac{\pi(t-T_{\rm cut})}{\Delta T}\right)}\right) & {\rm otherwise}
        \end{array}\right.
\end{equation}
with $T_{\rm cut} = t_c - 6.7\times 10^3 \, s$ and $\Delta T= 5\times 10^4\, s$. The same window was applied to the frequency domain waveforms using the time-frequency mapping $t(f)$. This choice of parameters removes the late inspiral, merger and ringdown, and reduces the signal-to-noise ratio to ${\rm SNR}=11.1$. Note that more rapid truncations with smaller $\Delta T$ result in unacceptable spectral leakage and large Gibbs oscillations in the frequency domain signals. 

\begin{figure}[htp]
\includegraphics[width=0.5\textwidth]{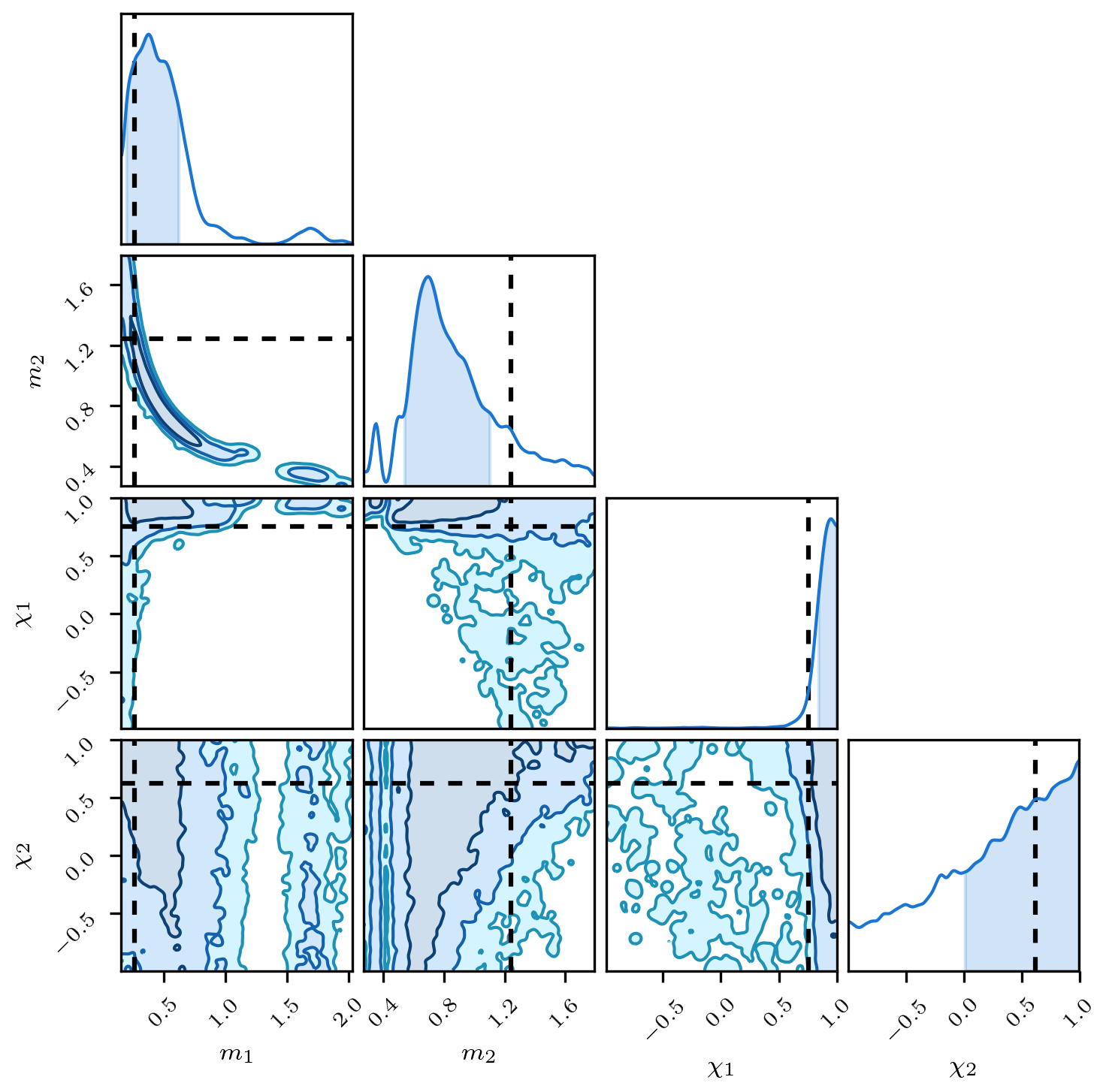} 
\caption{\label{fig:ms} Posterior distributions for the masses and spins using the truncated {\em Radler} data. The truncated signal has ${\rm SNR}=11.1$. The two-dimensional distribution for the masses exhibits the characteristic ``banana'' shape along the line of constant chirp mass. The instrument noise pushes the masses away from their true values.}
\end{figure}

\begin{figure}[htp]
\includegraphics[width=0.5\textwidth]{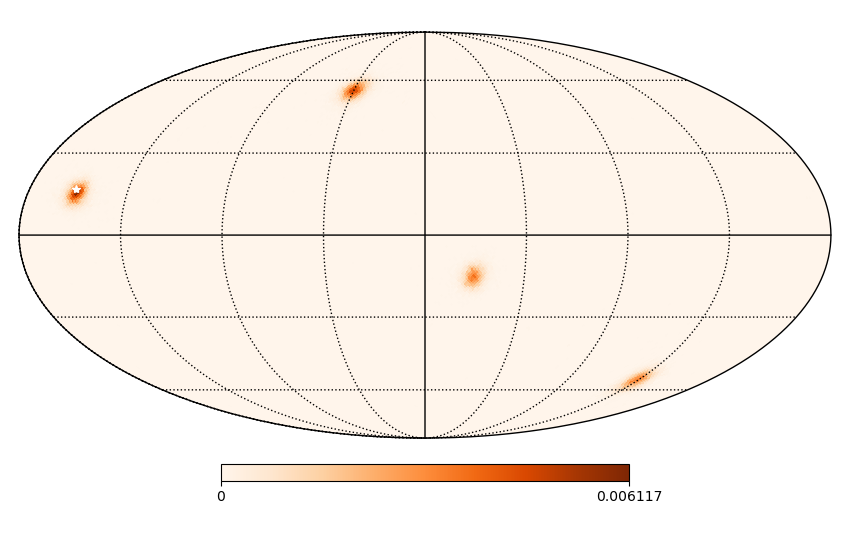} 
\caption{\label{fig:sky11} Sky map found using the truncated pre-merger {\em Radler} data. The truncated signal has ${\rm SNR}=11.1$.  The sky map exhibits the typical multi-modality of a short duration burst signal. The true source location is indicated by a white star.}
\end{figure}

Posterior distributions for the masses and spins using the truncated pre-merger signal are shown in Figure~\ref{fig:ms}. The mass distribution follows a line of constant chirp mass, as expected for the inspiral-only portion of the signal. The spin of the more massive system is already quite well constrained. The instrument noise pushes the masses and spins away from their true values. The posterior distribution for the ecliptic latitude and longitude of the truncated signal are shown in Figure~\ref{fig:sky11}. The short duration, burst-like nature of the signal results in four distinct modes. The $90\%$ credible interval covers 630 square degrees.

\begin{figure}[htp]
\includegraphics[width=0.5\textwidth]{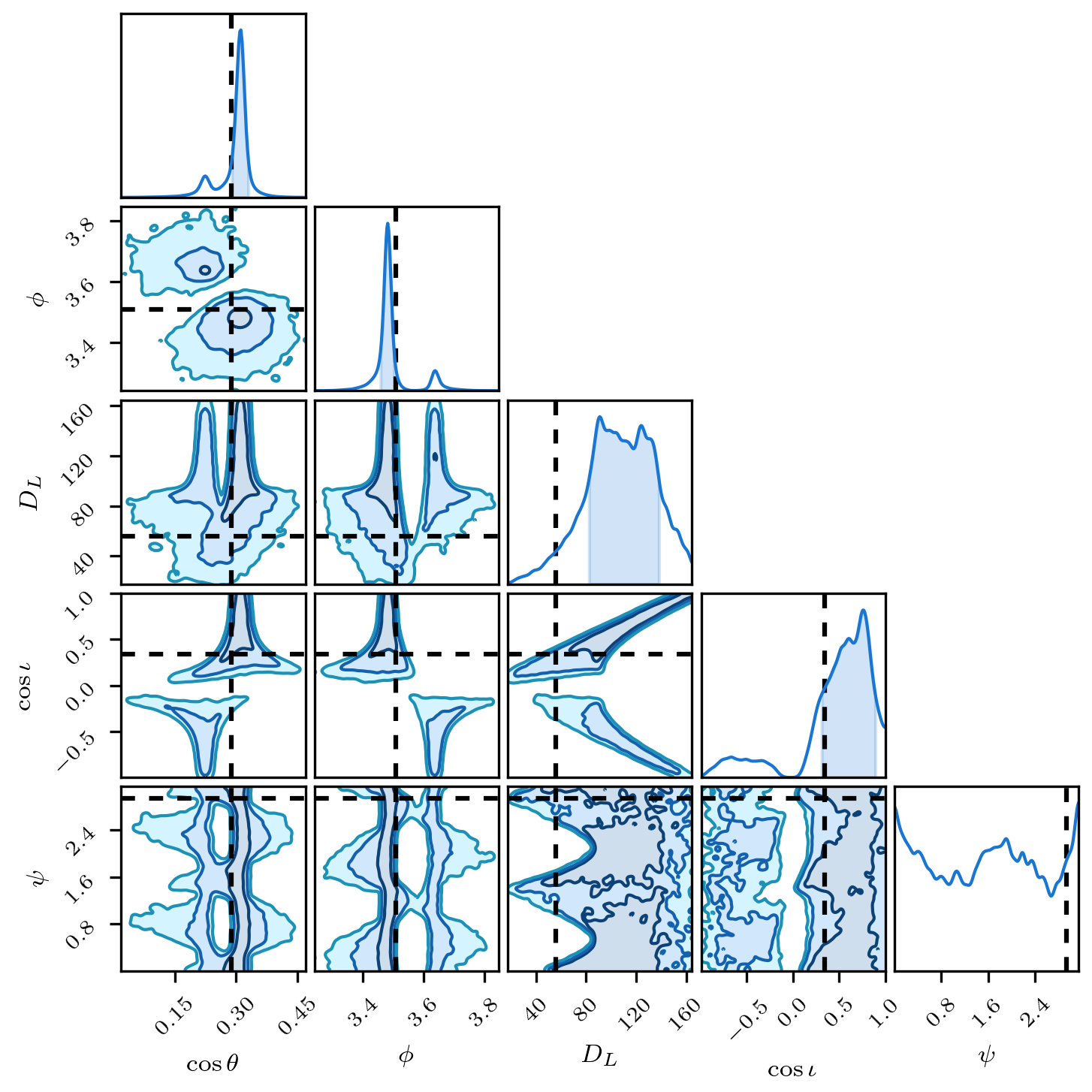} 
\caption{\label{fig:ex} Slices through the posterior distribution for the masses, spins, merger phase and merger time. The true values are indicated by dashed lines. Here the offsets from the true values is mostly due to projection effects and not noise - the peak of the posterior in the full $D=11$ dimensions can align with the true values but appear offset in the one and two dimensional projections shown here.}
\end{figure}

Including the merger and ringdown boosts the signal-to-noise ratio to ${\rm SNR}=405$ and reduces the $90\%$ credible interval for the sky location to 47 square degrees. Of the original four modes for the sky location only one survives, but the surviving mode splits into two closely space modes as seen in Figure~\ref{fig:ex}. The instrument noise moves the extrinsic parameters away from their true values, but the effect on the projected posterior distribution is small compared to the apparent displacements caused by the projections. To see this, compare the noise-free distributions shown in Figure~\ref{fig:comp_ex} to the analysis with noise shown in Figure~\ref{fig:ex}. The luminosity distance is significantly impacted by these projection effects.

\begin{figure}[htp]
\includegraphics[width=0.5\textwidth]{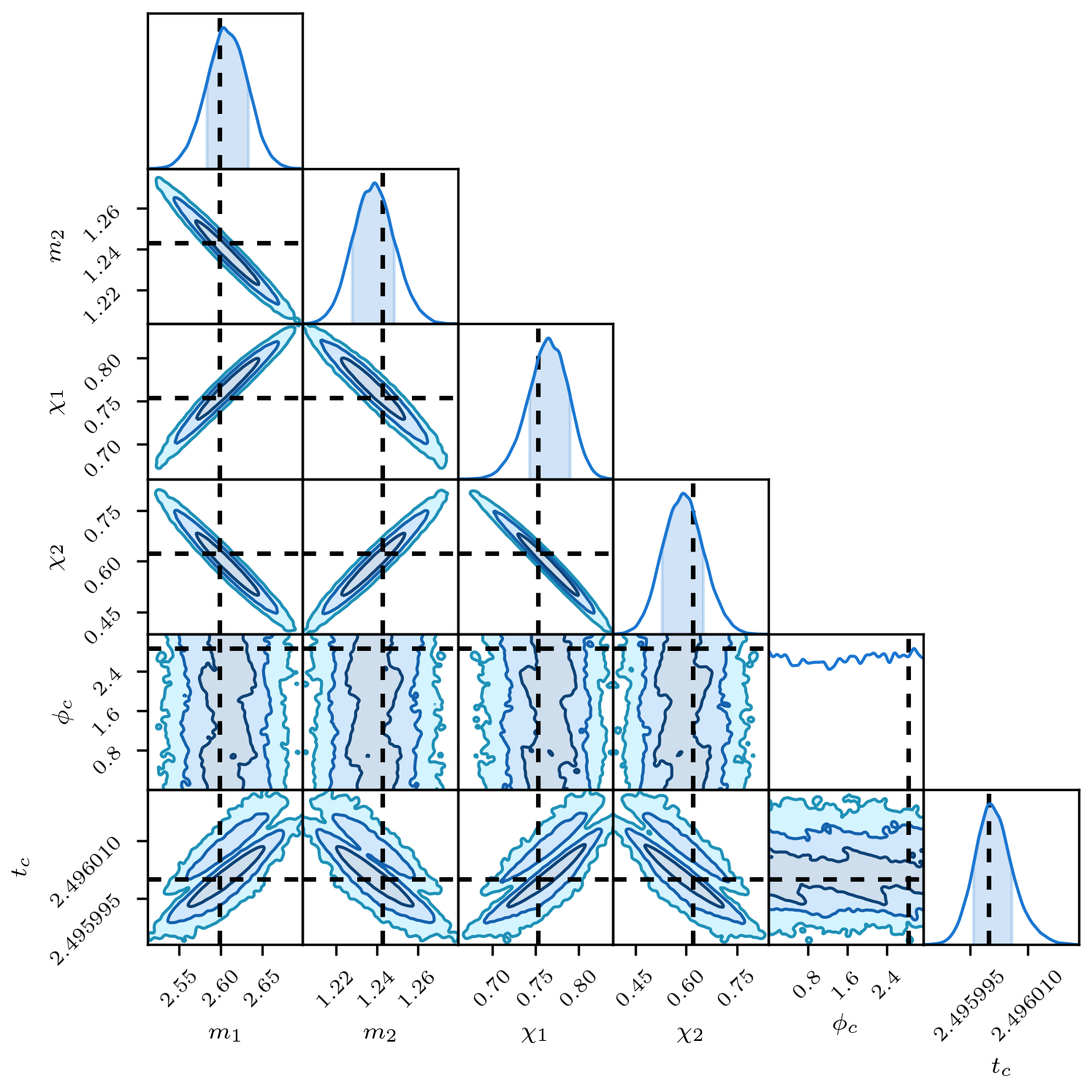} 
\caption{\label{fig:in} Slices through the posterior distribution for the masses, spins, merger phase and merger time. The true values are indicated by dashed lines. The peak of the posterior distribution is offset from the true values by the noise. The offsets are relatively small for the particular noise realization used in the {\em Radler} data set.}
\end{figure}

The parameters that directly enter the phase, shown in Figure~\ref{fig:in} are shifted slightly by including the instrument noise. The smallness of the shifts is just luck of the draw - repeating the analysis with different noise realizations yielded larger shifts on average.

\section{Discussion}

We have presented an early prototype for detecting and characterizing massive black holes with LISA. A key feature of the methods we have developed is speed. All the analyses were conducted on a 2016 vintage laptop, with the search stage taking tens of minutes and the characterization stage taking a few hours. The simulated data we analyzed is much simpler than what we expect the real LISA data will look like. In reality we will have to contend with non-stationary and non-Gaussian noise, data gaps, more complex waveforms and multiple overlapping signals. We aim to tackle each of these complications in future work that builds on the foundation we have laid here.

\section*{Acknowledgments}
We appreciate the support of the NASA LISA foundation Science Grant 80NSSC19K0320. This work was initiated while NJC was on sabbatical at the Observatoire de la C\^{o}te d'Azur, kindly hosted by Nelson Christensen, and supported in part by the Centre National d'\'{E}tudes Spatiales. We have benefited from many exhanges with Stanislav Babak and Tyson Littenberg.

\appendix

\section{Noise model}
\label{a:noise}
    
The noise spectral density in the fractional-frequency $A,E$ TDI channels is modeled as
\begin{eqnarray}
&& S_n(f) = \frac{64}{3 L^2} \left(\frac{f}{f_*} \sin\left( \frac{f}{f_*} \right)\right)^2 \left[  \left(2+\cos\left( \frac{f}{f_*} \right)\right) S_{\rm ps} \right. \nonumber \\
&& \hspace*{0.5in} + \left( 6 + 4 \cos\left( \frac{f}{f_*} \right) + 2\cos\left( \frac{2f}{f_*} \right) \right)   \nonumber \\
&& \hspace*{0.5in} \left. \times \frac{S_{\rm acc}}{(2\pi f)^4}\left(1+16\left(\frac{10^{-4}}{f}\right)^2\right) \right] \, .
\end{eqnarray}
with position noise level $S_{\rm ps} = 2.25\times 10^{-22}\,  {\rm m^2} {\rm Hz}^{-1}$ and acceleration noise level $S_{\rm acc} = 9\times 10^{-30}\, {\rm m^2} {\rm s}^{-4} {\rm Hz}^{-1}$.

\section{Frequency Domain Instrument Response}
\label{a:response}

The LISA constellation cartwheels around the Sun resulting in a time dependent instrument response function. Since the likelihood is computed in the frequency domain, and the PhenomD waveform is already expressed in the frequency domain, it is most efficient to map the frequency to time and compute the response directly in the frequency domain, as was first proposed by Cutler~\cite{Cutler:1997ta}. The mapping is given by
\begin{equation}\label{tf}
t(f) = \frac{1}{2\pi} \frac{d\Phi(f)}{df} + t_c \, .
\end{equation}
The time-frequency mapping is computed by taking numerical derivatives of the PhenomD phase function. In the expressions below it is understood that the time $t$ is mapped to the frequency $f$ by $t(f)$. 

The fractional frequency shift imparted by a gravitational wave to the laser light propagating along one spacecraft to another is given in equation (B11) of Ref.~\cite{Rubbo:2003ap}, which can be written in full generality as
\begin{equation}
\frac{\delta \nu_{ij}(t)}{\nu_0} =  \frac{ \hat{r}_{ij} \otimes \hat{r}_{ij}}{2(1 - \hat{k}\cdot \hat{r}_{ij})} : \left({\bf h}(t,x_j)-{\bf h}(t-L,x_i)\right) \, .
\end{equation}
This expression describes the Doppler shift of the laser light in going from the spacecraft at ${\bf x}_i$ to the spacecraft at ${\bf x}_j$, arriving at Barycenter time $t$. Here $\hat{r}_{ij} = ({\bf x}_j-{\bf x}_i)/L$, where $L$ is the armlength, which we are assuming is constant. The gravitational wave is propagating in the $\hat{k}$ direction with surfaces of constant phase given by $\xi = t- \hat{k}\cdot {\bf x}$. The full TDI response is formed out of a linear combination of one-arm Doppler shits along various arms at various times. In order to be able to add together these Doppler shifts in a consistent way, it is helpful to reference all of expression to the gravitational wave signal at the center of the constellation, given by ${\bf x}_0 = ({\bf x}_1 +  {\bf x}_2 + {\bf x}_3)/3$. Working in the rigid-adiabatic approximation~\cite{Rubbo:2003ap} for a GW signal with instantaneous frequency $f$ we have
\begin{eqnarray}
&&\frac{\delta \nu_{ij}(t)}{\nu_0} =  i (\hat{r}_{ij} \otimes \hat{r}_{ij}):{\bf h}(t,{\bf x}_0)\left[{\rm sinc}\left( \frac{f}{2 f_*}(1-\hat{k}\cdot \hat{r}_{ij})\right) \right.  \nonumber \\
&& \hspace*{1in} \left. \times e^{-i \frac{f}{2 f_*} (1 + \hat{k}\cdot \hat{r}_{ij} - \frac{2}{\sqrt{3}} \hat{k}\cdot \hat{r}_{i0})} \right]
\end{eqnarray}
where $f_* = 1/(2 \pi L)$ is the transfer frequency. The gravitational wave signal is given by 
\begin{equation}
{\bf h}(\xi) = h_+(\xi) \boldsymbol{\epsilon}^+ +  h_\times(\xi) \boldsymbol{\epsilon}^\times \, .
\end{equation}
For the leading order 22-mode of a non-precessing circular binary we have
\begin{equation}
{\bf h}(\xi) = h(\xi)\left( A_+ \boldsymbol{\epsilon}^+ +  i A_\times \boldsymbol{\epsilon}^\times \right)\, ,
\end{equation}
where
\begin{equation}
A_+  =\frac{1+\cos^2\iota}{2}\, , \quad A_\times = -\cos\iota \, ,
\end{equation}
and $\iota$ is the inclination of the binary orbit. 

In the rigid adiabatic approximation the time delay interferometry introduces an overall transfer function given by
\begin{equation}
{\cal T} = (1-e^{-4\pi i f L}) = 2 i e^{-i \frac{f}{f_*}} \sin(f/f_*) \, .
\end{equation}
Putting all the pieces together, the X channel TDI variable extracted from vertex 1 is given by
\begin{eqnarray}
&&X(t) = -\frac{f}{f_*} e^{-i \frac{f}{f_*}} \sin\left(\frac{f}{f_*}\right)\left[ (\hat{r}_{12} \otimes \hat{r}_{12}) {\cal T}_{12}(t)  \right.  \nonumber \\
&&\hspace*{0.5in} \left.  -   (\hat{r}_{13} \otimes \hat{r}_{13}) {\cal T}_{13}(t)\right]:{\bf h}(t,{\bf x}_0)
\end{eqnarray}
where
\begin{eqnarray}
&&{\cal T}_{ij}(t) = {\rm sinc}\left( \frac{f}{2 f_*} \left( 1 - \hat{k}\cdot \hat{r}_{ij}\right) \right)e^{-i \frac{f}{2 f_*}(3 +  \hat{k}\cdot \hat{r}_{ij} - \frac{2}{\sqrt{3}}  \hat{k}\cdot \hat{r}_{i0})} \nonumber \\
&& \hspace*{0.1in} + {\rm sinc}\left( \frac{f}{2 f_*} \left( 1 + \hat{k}\cdot \hat{r}_{ij}\right) \right)e^{-i \frac{f}{2 f_*}(1 +  \hat{k}\cdot \hat{r}_{ij} - \frac{2}{\sqrt{3}} \hat{k}\cdot \hat{r}_{i0})}.
\end{eqnarray}
If we define $d_{ij}^{+,\times} = (\hat{r}_{ij} \otimes \hat{r}_{ij}):{\epsilon}^{+,\times}$, the 22-mode response can be written as
\begin{equation}
X(t) =\left[F_X^+(t) A_+ + i F_X^\times(t) A_\times\right] h(t,{\bf x}_0) \, ,
\end{equation}
where the complex antenna patterns are given by
\begin{equation}
F_X^{+,\times}  = -\frac{f}{f_*} e^{-i \frac{f}{f_*}} \sin\left(\frac{f}{f_*}\right) \left(d_{12}^{+,\times}\,   {\cal T}_{12} - d_{13}^{+,\times}\, {\cal T}_{13} \right) \, .
\end{equation}
Expressions for the $Y$ and $Z$ channels follow by cyclic permutation of the labels $(1,2,3)$ in the expression for $X$. The signal orthogonal $A,E,T$ can be formed out of linear combinations of $X,Y,Z$:
\begin{eqnarray}
A &=& \frac{1}{3}\left( 2 X - Y - Z\right) \nonumber \\
E &=&  \frac{1}{\sqrt{3}}\left( Z - Y\right) \nonumber \\
T &=& \frac{1}{3}\left( X + Y + Z\right) \, .
\end{eqnarray}
For the $A,E$ channels used in the analysis, the miss-match between the noiseless {\em Radler} data and the frequency domain rigid adiabatic waveforms was
${\rm MM}_A = 3.6\times 10^{-7}$ for the $A$ channel and ${\rm MM}_E = 7.7\times 10^{-7}$ for the $E$ channel. These systematic mis-matches are well below
the expected statistical mis-matches due to noise, ${\rm E}[{\rm MM}] = (D-1)/(2 {\rm SNR}^2)$, for a signal with dimension $D=11$ and ${\rm SNR}_A = 360.9$, ${\rm SNR}_E = 184.9$.

\section{Heterodyned Likelihood}
\label{a:hlike}

The log-likelihood in Gaussian noise is given by
\begin{equation}
\log L = \frac{1}{2}(d-h|d-h) -\frac{1}{2} \int 2\pi S_n(f) \, df \, ,
\end{equation}
where $d$ is the data, $h$ is the waveform model, $S_n(f)$ is the noise spectral density and the notation $(a|b)$ indicates that usual noise weighted inner product. In an effort to reduce clutter in the notation we will suppress the sum over channels in what follows.

The idea behind the heterodyned likelihood~\cite{Cornish:2010kf}  is that given a good reference model $\bar h$, such as the maximum likelihood waveform found in the search phase, the likelihood for waveforms ``close'' to $h$ can be computed by heterodyning the residual, $\bar{r} = d-\bar h$, against the signal $\bar h$.
This results in a likelihood that can be computed very cheaply using a coarse spline interpolation of the amplitude and phase. In an MCMC, any parameters that are in the central 99.9+\% of the posterior will generate waveforms that are close enough to the reference waveform for the heterodyned likelihood to be used. In fact, the heterodyned likelihood itself is {\em exact}. It is the approximations used to make the computation fast that introduce error, and the size of the error can be controlled by how many terms are kept in the splines and FFTs used to speed up the evaluation.

In contrast to the reduced order quadrature method for accelerating the likelihood evaluation, the heterodyned likelihood is able to accommodate updates to the noise model. While the noise model was held fixed in the current application, noise updates are included here for completeness. Given a reference waveform $\bar h$ and noise model $\bar{S}_n(f)$. The second term in the likelihood can be computed directly using a spline integration. The first term in the likelihood requires more attention:
\begin{eqnarray}
&&(d-h|d-h)  = 4 \int \frac{(\bar{r} + \Delta h)(\bar{r} + \Delta h)^*}{S_n(f)} df  \nonumber \\
&& \hspace*{0.2in} = 4 \int \frac{ \Delta h \Delta h^*  + (\bar{r}\Delta h^* + \bar{r}^* \Delta h) + \bar{r} \bar{r}^*}{S_n(f)} df \, ,
\end{eqnarray}
where $ \Delta h  = \bar h - h$. Writing $h={\cal A}(f) e^{i\Phi(f)}$, and similarly for $\bar h$, we have
\begin{equation}\label{dh}
\int \frac{ \Delta h \Delta h^* }{S_n(f)} df = \int \frac{ \bar{{\cal A}}^2(f) + {\cal A}^2(f) - 2  \bar{{\cal A}}(f){\cal A}(f)\cos\Delta \Phi(f)}{S_n(f)} df \, .
\end{equation}
with $\Delta \Phi(f) = \bar{\Phi}(f) - \Phi(f)$. Only the phase difference appears here since $\bar h$ naturally heterodynes $h$. This expression is exact. So long as $h$ is close to $\bar h$, all the terms in the integrand are slowly varying and can be evaluated using a coarse spline interpolation. Next we have
\begin{equation}
\int \frac{ \bar{r}\Delta h^* + \bar{r}^* \Delta h }{S_n(f)} df = \int   (\bar{r}_w \Delta h^*_w + \bar{r}^*_w \Delta h_w) df \, ,
\end{equation}
where
\begin{equation}
\bar{r}_w = \frac{\bar{r} \,e^{-i \bar{\Phi}(f)}}{\bar{S}^{1/2}_n(f)}
\end{equation}
is the whitened reference residual heterodyned by the reference phase and
\begin{equation}
\Delta h_w =\frac{ \left(\bar{\cal A}(f)  - {\cal A}(f) e^{-i\Delta \Phi(f)} \right)\bar{S}^{1/2}_n(f)}{S_n(f)} 
\end{equation}
 is the heterodyned and whitened difference in the waveforms. Writing
\begin{equation}
\bar{r}_w(f) = \int e^{2\pi i \tau}  \bar{r}_w(\tau) \, d\tau \, ,
\end{equation}
and similarly for $\Delta h_w(f)$ we have
\begin{equation}
\int \frac{ \bar{r}\Delta h^* + \bar{r}^* \Delta h }{S_n(f)} df = \int   (\bar{r}_w(\tau)\Delta h^*_w(\tau) + \bar{r}^*_w(\tau) \Delta h_w(\tau)) d\tau \, .
\end{equation}
This expression is exact. It can be approximated by using a FFT to compute the Fourier transform and using a restricted range for the $\tau$ parameter. Note that the more expensive to compute $\bar{r}_w(\tau)$ can be evaluated once and stored.The final term can be handled in a similar fashion:
\begin{equation}
\int \frac{  \bar{r} \bar{r}^* }{S_n(f)} df = \int  R(f) S(f) df =  \int  R^*(\tau) S(\tau) d\tau\, ,
\end{equation}
where $R(f) =  \bar{r} \bar{r}^*/\bar{S}_n(f)$, $S(f)= \bar{S}_n(f)/{S}_n(f)$ and $R(\tau)$ and $S(\tau)$ are their Fourier transforms.
As with $\bar{r}_w(\tau)$, the expensive to compute $R(\tau)$ can be evaluated once and stored.

The reference integrals $\bar{r}_w(\tau)$ and $R(\tau)$ are calculated at the full sample cadence of the data, while the slowly varying terms such as (\ref{dh}) are computed on a coarse spline in frequency. In the LISA context we want to ensure that the orbital motion of the constellation is adequately sampled, so we use the leading post-Newtonian expression for $\dot f$ to set the frequency spacing $df$:
\begin{equation}
df =  \dot f dT =  (8\pi)^{8/3} \frac{3}{40} {\cal M}^{5/3} dT \, .
\end{equation}
Setting $dT = 3\times10^5$ seconds yields rough 100 hundred samples per year during the early inspiral. To ensure that the dynamic frequency spacing is never to fine or too coarse we set $df_{\rm min} = 1/T_{\rm obs}$ and $df_{\rm max} = f_{\rm ring}/100$, where $f_{\rm ring}$ is the ringdown frequency. With these choices, the frequency stencil typically has between 100 and 500 points for a one year data set.

For the discrete FFT used to compute $\Delta h_w(\tau)$ we settled on $N=4096$ points in a trade-off between speed accuracy. This choice delivered at accuracy of order $\pm 0.3$ for the $(n|\Delta h)$ term, with an evaluation time of 5 ms on a single 2.6 GHz core. The accuracy should be compared to the expected value and variance for this term, ${\rm E}[(n|\Delta h)] \simeq {\rm E}[(n|h_{,i} (n|h_{,j})\Gamma^{ij})]= D$, and ${\rm Var}[(n|\Delta h)] = D$. The standard deviation of the
$(n|\Delta h)$ term, $\sigma_{(n|\Delta h)} = \sqrt{D} = 3.3$, is much larger than the numerical error.

\bibliography{refs}

\end{document}